\documentclass{webofc}

\usepackage{hyperref}

\usepackage[varg]{txfonts}   
\usepackage{tikz}
\usepackage{bm}
\usepackage{graphicx}   
\usepackage{amsmath,amssymb,mathrsfs}  
\usepackage[utf8]{inputenc}
\usepackage{color}

\def\mn{{\mu\nu}}

\begin{document}
\title{Colliding poles with colliding nuclei
}

\author{
        \firstname{Alexander} \lastname{Soloviev}\inst{1}\fnsep\thanks{\email{alexander.soloviev@tuwien.ac.at}} 
}

\institute{Institut f{\"u}r Theoretische Physik, Technische Universit{\"a}t Wien,\\
Wiedner Hauptstr. 8-10, A-1040 Vienna, Austria
          }

\abstract{%
In these proceedings, I will discuss collisions of poles in the complex plane as a signature of phase transitions for theories relevant to the quark gluon plasma. I will begin with an illustrative example, namely the chiral phase transition, which can be characterized by colliding poles as a function of temperature. Then, recognizing the interplay between weak and strong coupling sectors in a typical collision, I will introduce a hybrid model with a weakly broken symmetry, which has a rich quasi-hydrodynamic phenomenological description where hydrodynamic and non-hydrodynamic poles are unified by a common dispersion relation. I will show that energy is transferred initially from the soft to the hard sector before irreversibly transferring back to the soft sector at late times, and that the model reproduces many features common to dissipative systems with a weakly broken symmetry including the k-gap.
}
\maketitle



\section{Introduction} \label{intro}

In the wake of heavy ion collisions, performed in experiments in the LHC at CERN and RHIC at BNL, a state of matter of liberated quarks and gluons, known as the quark-gluon plasma (QGP), is formed \cite{STAR:2005gfr}. There are three common effective descriptions used to study the QGP, which have different regions of applicability within the evolution of a typical heavy ion collision. At early times, kinetic theory captures the weakly coupled dynamics \cite{Arnold:2002zm}. At later times, as the system undergoes hydrodynamization, strongly coupled holography \cite{Heller:2016gbp} and relativistic hydrodynamics \cite{Teaney:2009qa} become increasingly more relevant.

 Following the discussion in \cite{Kurkela:2019kip}, we can examine the analytic structure of the mentioned theories. Each theory mentioned has in common a hydrodynamic pole, but other non-analytic structures are radically different. For instance, kinetic theory in the relaxation time approximation typically has a branch cut between $-k$ and $k$ at $-i/\tau$ where $\tau$ is the relaxation time \cite{Kurkela:2017xis}. Holographic theories exhibit a so-called "Christmas tree" structure of an infinite number of poles with increasing real and imaginary values in the lower half plane \cite{Grozdanov:2018gfx}. Hydrodynamic theories such as M\"uller-Israel-Stewart (MIS) \cite{Muller:1967zza,Israel:1979wp} have a non-propagating dissipative pole. 
One is left naturally to ask the question: what is the analytic structure and, ultimately, the microscopic structure of the QGP?

Here, we will aim to partly answer the above question by first underlining in Sec.~\ref{sec:oh4} that the chiral phase transition in the $O(4)$ model \cite{Rajagopal:1992qz, Son:1999pa,Son:2002ci}, a characteristic feature present in a typical heavy ion collision, can be understood via a collision of poles in the complex plane, as the system moves from diffusive to propagating degrees of freedom. Then in Sec.~\ref{sec:semi}, we will demonstrate a similar collision of poles in a theory with weak/strong coupling dynamics via a quasinormal mode analysis of a semiholographic model with a boundary scalar field.

\section{Colliding poles near the O(4) critical point}\label{sec:oh4}

\begin{figure}[htp]
\centering
\includegraphics[width=0.5\textwidth]{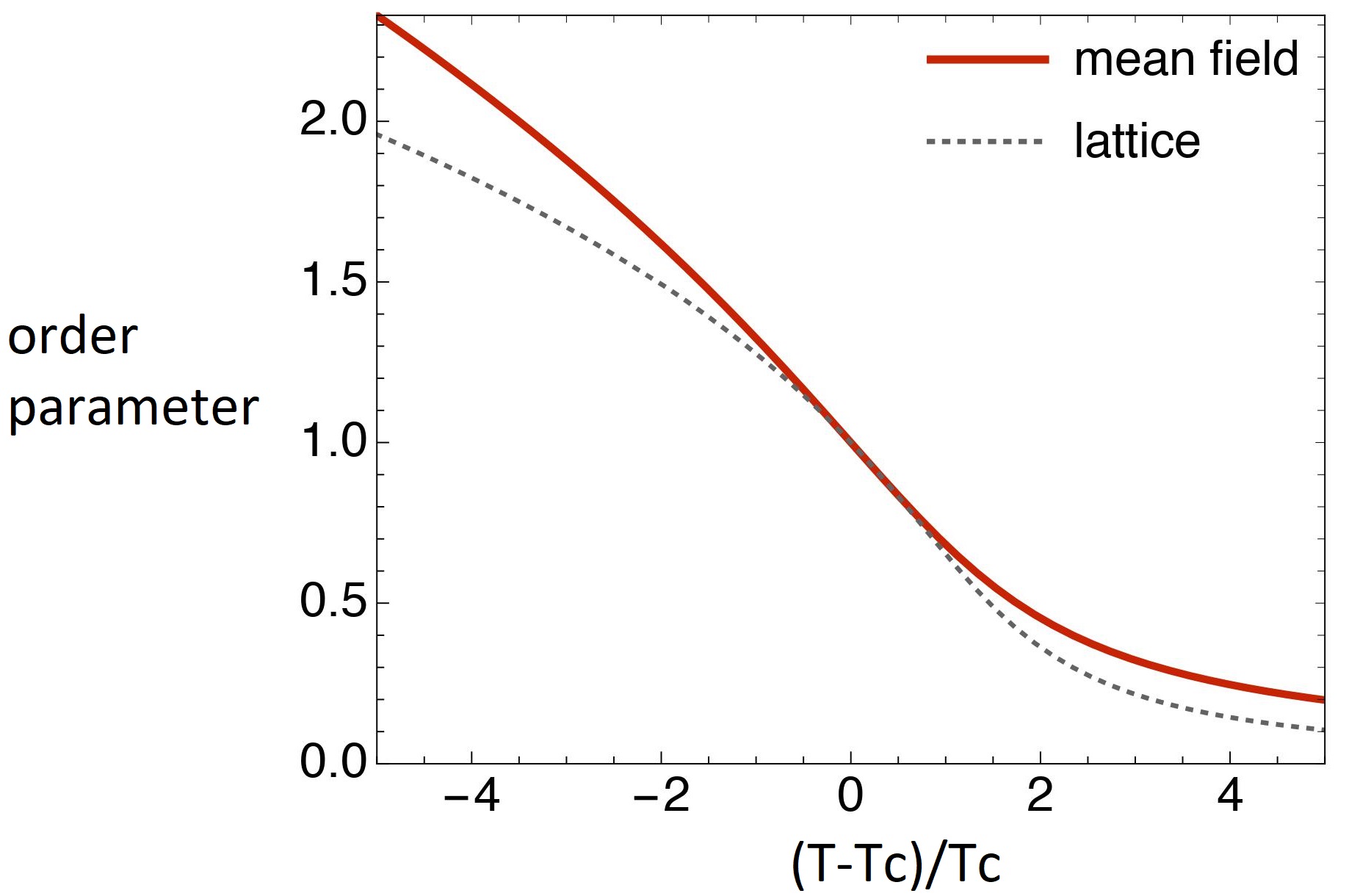}
\caption{Comparison of mean field order parameter to the lattice in \cite{Engels:2009tv,	Engels:2011km}.}\label{fig:mean}
\end{figure}

Here, we will discuss reinterpret the chiral phase transition, explored via the $O(4)$ model in \cite{Grossi:2020ezz,Grossi:2021gqi, Florio:2021jlx,Soloviev:2021syx}, as a collision of poles in the complex plane.
We describe the $O(4)$ physics via the effective Hamiltonian 
\begin{align}\label{ham}
\mathcal{H} =\int_x &p(T)+\frac{\chi_0}{4}\mu_{ab}^2 
-\frac{1}{2}\Delta^{\mu\nu}D_\mu\phi_a D_\nu\phi_a+\frac{1}{2}m^2_0 (T-T_c) \phi_a\phi_a+\frac{\lambda}{4} (\phi_a\phi_a)^2-H_a\phi_a
\end{align}
where we denote the $O(4)$ vector via $\phi_a=(\sigma, \varphi_i)$, $\Delta^\mn=u^\mu u^\nu +g^\mn,$ $\mu_{ab}$ is the $O(4)$ chemical potential, $\chi_0$ is the static susceptibility and $H_a$ is the explicit symmetry breaking term. Note that the mass term, $m^2=m_0^2(T-T_c)$, changes sign in the vicinity of the critical temperature. 
Choosing $H_a=(H,0,0,0),$ we will work in the mean field limit, which is found by minimizing the potential in \eqref{ham} to find
\begin{align}
m^2_0(T-T_c) \sigma + \lambda \sigma^3-H=0,
\end{align}
which determines the mean field value of the condensate as a function of temperature, $\sigma(T)$, with its behavior shown in Fig.~\ref{fig:mean}.

\begin{figure}[htp]
  \begin{center}
 \includegraphics[width=0.48\textwidth]{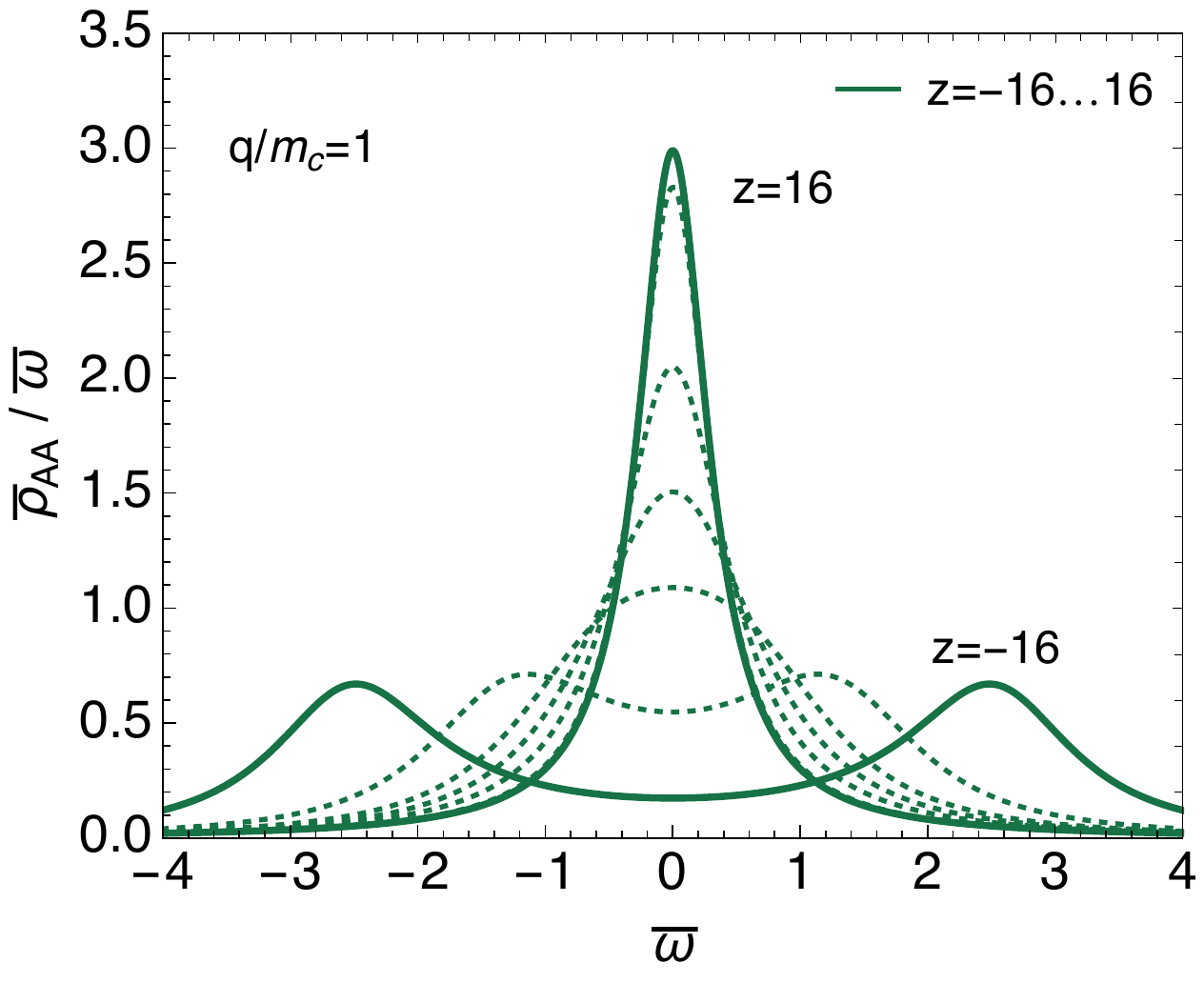}
\includegraphics[width=0.5\textwidth]{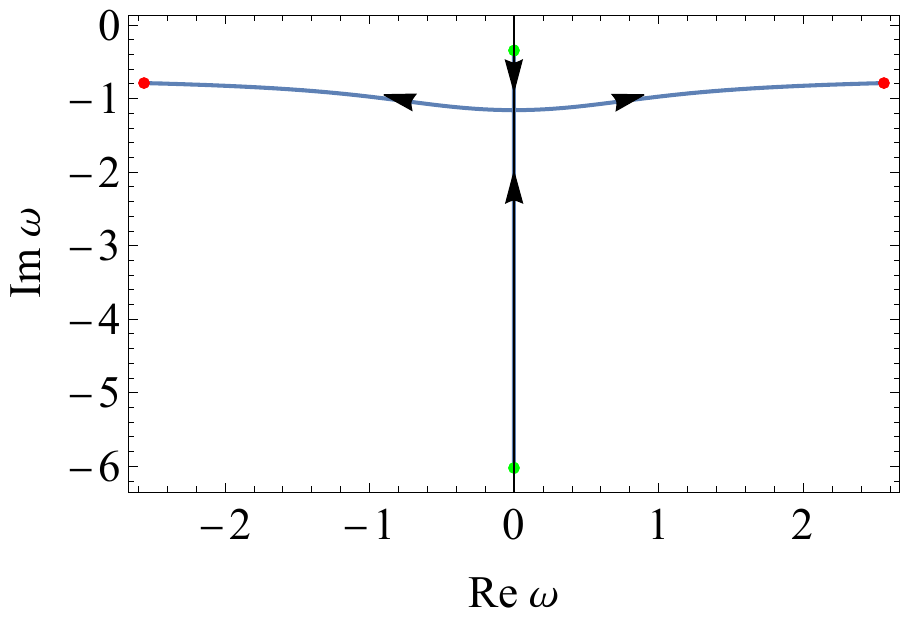}
    \end{center}
\caption{Two different ways to see the $O(4)$ phase transition via the spectral density. Left: The spectral density over frequency for different $z\sim T-T_c$. For the choice of parameters used in this plot, see the discussion in \cite{Grossi:2021gqi}. The diffusive peak in the unbroken phase $z\gg 0$ melts to two peaks in the broken phase, $z\ll 0$, representing the propagating pions. Right: the pole structure of the spectral density as a function of temperature. The green points represent $z=16$, i.e. the high temperature, unbroken phase, while the arrows indicate the decrease in temperature until the red point $z=-16.$ Like in the left plot, the diffusive poles on the imaginary axis melt down to two propagating poles. }
\label{fig:prop}
\end{figure}

Computing the ideal equations of motion from \eqref{ham}, including dissipation and linearizing around mean field leads to the following coupled equations for the pions and chemical potential:
\begin{align}
{\partial_t\varphi}&=
-{{\mu}_A}
+{\Gamma(\nabla^2 -m^2)\varphi}
\\
{\partial_t \mu_A}&= {v^2(-\nabla^2+m^2)\varphi}
+{D_0 \nabla^2 \mu_A}
\end{align}
We do not consider the dynamics of the condensate, which decouple from the pions and the chemical potential in the linearized limit. From the above, we can determine the symmetrized propagators $G_{\rm sym}=T(G_R -G_A)/(i\omega)$. Here, we will focus on the
 the spectral density for the axial charge density-density correlator, which is related to the $\varphi\varphi$ symmetrized propagator via
\begin{align}
\rho_{AA}=\frac{\omega}{T (\chi_0 \omega_k)^2}G^{\varphi\varphi}_{\rm sym},
\end{align}
where $\omega_k^2=v^2(k^2+m^2).$
The behavior of this propagator is best understood by considering Fig.~\ref{fig:prop}. In the left panel, reproduced from \cite{Grossi:2021gqi}, the spectral function is plotted as a function of frequency for various $z\sim (T-T_c)/T_c$. At high temperatures, the system is in the unbroken phase, characterized by a diffusive peak that we can interpret as the diffusion of quarks in the QGP. As the temperature decreases, the system eventually goes through the phase transition, with the spectral function exhibiting two well-separated peaks that we can interpret as pions. 

The right panel presents another way to view the same information in the left panel, but from the perspective of the pole structure of the spectral function. Starting from the green diffusive poles, representing the unbroken phase with $T\gg T_c$, the direction of decreasing temperature is indicated by the black arrows. Eventually the diffusive from the two poles collide, indicating that the system is undergoing a phase transition. From the collision point, further decreasing the temperature leads to two well separated poles, namely the pion modes. 

In this way, the collision of poles as a function of temperature in the complex plane indicates a phase transition, a feature that should be studied in understanding the analytic structure of the QGP.

\section{Colliding poles in holography}\label{sec:semi}

We now turn our attention to understanding the analytic structure of the QGP from another perspective. Although strong and weak coupling dynamics are often studied separately, a complete description of the QGP would entail understanding the interplay between the two sectors. One approach in this line of reasoning is known as semiholography \cite{Faulkner:2010tq, Iancu:2014ava, Mukhopadhyay:2015smb, Banerjee:2017ozx,Kurkela:2018dku,Ecker:2018ucc, Mitra:2020mei,Mitra:2022uhv}, a framework which self-consistently mixes {holography in the infrared} with perturbative degrees of freedom in the ultraviolet via dynamical boundary fields. 

\begin{figure}
\center{
\includegraphics[width=0.65\linewidth]{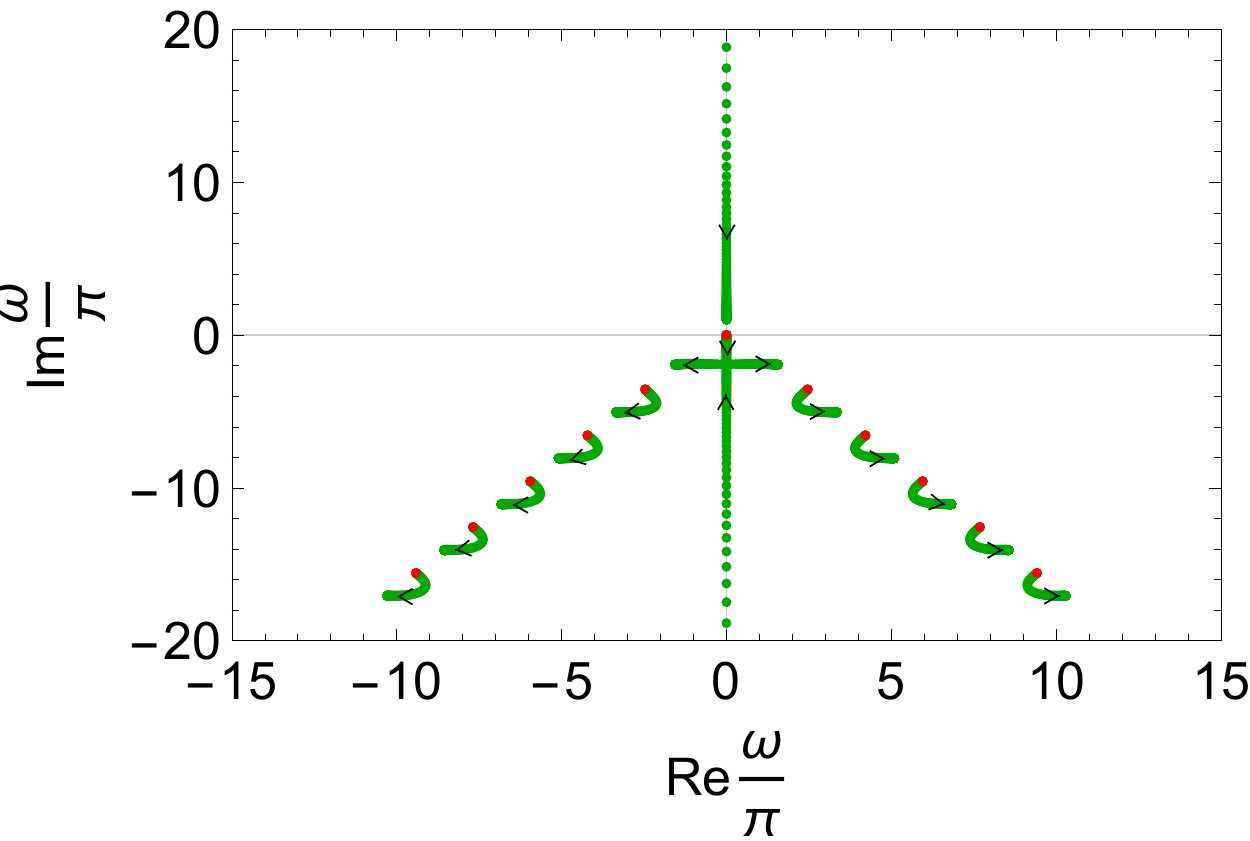}
}\label{fig:homo-qnm}
\caption{
Homogeneous quasinormal modes with arrows indicating the direction of increasing coupling, $\beta$. Red points denote the $\beta=0$ case, which is the typical "Christmas tree" pattern of the QNMs. A collision of poles indicates the breaking of the softly broken global symmetry.}
\end{figure}

As a quick recap we first discuss a simple example of quasinormal modes (QNMs) found in holography, before moving onto the example of semiholography. QNMs describe dissipation of linearized perturbations around equilibrium solutions, i.e. the ring down of modes. As an example, consider the equation of motion for a bulk massless scalar field: 
\begin{align}
\nabla_M \nabla^M \Phi =0,
\end{align}
living in the Eddington-Finkelstein Schwarzschild-AdS$_4$ background:
\begin{align}
ds^2=-\frac{L^2}{r^2}(1-Mr^3)dt^2-2 \frac{L^2}{r^2}dtdr+\frac{L^2}{r^2}(dx^2+dy^2).
\end{align}
We can decompose the scalar field into Fourier modes $\Phi(r,x^\mu)\rightarrow e^{-ix\cdot k}f(r,k_\mu)$, which leads to the linear equation
\begin{align} 
0&=(Mr^3-1)f^{\prime\prime}+\frac{Mr^3+2-2ir \omega}{r}f^{\prime}+\frac{k^2r+2i\omega}{r}f
\end{align}
Computing the poles gives rise to the typical "Christmas tree" structure found in holography \cite{Horowitz:1999jd,Kovtun:2005ev,Berti:2009kk}, see the red points of Fig.~\ref{fig:homo-qnm}.

We now add an additional dynamic scalar field at the boundary of the holographic theory described above and follow the discussion in \cite{Mondkar:2021qsf}. 
The semiholographic action reads
\begin{align}
S={W_{\rm CFT}}[{h(x)}=-\beta {\chi}] -\frac{1}{2}\int d^3 x\; { \partial_\mu \chi \partial^\mu \chi},
\end{align}
where $W_{\rm CFT}$ is the generating functional of the conformal field theory (dual to the gravitational theory), sourced by the boundary scalar, $\chi$, with $\beta$ governing the coupling between the two sectors.
The equation of motion for the scalar field is 
\begin{align}
{\eta^{\mu\nu}\partial_\mu \partial_\nu\chi}=\beta {\frac{\delta W_{\rm CFT}}{\delta h}}=\beta {\mathcal{H}},
\end{align}
where $\mathcal{H}$ is the vacuum expectation value of the bulk scalar field, $\Phi.$
The equations of motion of the dual gravity theory to the CFT are
\begin{align}
{R_{MN}-\frac{1}{2}R G_{MN}-3G_{MN}}&={\kappa (\nabla_M \Phi \nabla_N\Phi-\frac{1}{2}G_{MN} (\nabla_P \Phi)^2 )},\\
{\nabla_M\nabla^M\Phi}&=0.
\end{align}
The bulk scalar field, ${\Phi}$, has the near boundary expansion: 
\begin{align}
{\Phi}={-\beta {\chi}}
{+\ldots +\frac{3}{\kappa} \mathcal{H} r^3+\ldots}.
\end{align}
Note that $\Phi$ is sourced by the dynamical boundary field, $\chi.$

The individual subsector Ward identities are
\begin{align*}
\partial_\mu {t^\mn_\chi}=-{\mathcal{H}\partial^\nu h}, \quad 
\partial_\mu {\mathcal{T}^\mn} ={\mathcal{H}\partial^\nu h},
\end{align*}
where $t^\mn_\chi$ is the boundary scalar stress tensor and $\mathcal{T}^\mn$ is the holographic stress tensor
The energy momentum tensor of the full system is conserved, $\partial_\mu T^\mn=\partial_\mu({t^\mn_\chi}+{\mathcal{T}^\mn})=0$.

We now study hybrid fluctuations of bulk dilaton ${\Phi}$ and the boundary scalar field ${\chi}$. In the case of homogeneous QNM,  we see in Fig.~\ref{fig:homo-qnm} that varying the coupling $\beta\sqrt{T}\rightarrow\infty$ moves the "Christmas tree" structure down into the complex plane and in the limit $\beta\sqrt{T}\rightarrow \infty,$ the system exhibits emergent conformality. Moreover, as we vary the coupling, there is a similar collision of poles on the imaginary axis as in the previous section. In this case, the diffusion-like poles become aligned on the "Christmas tree" structure for large coupling.

  \begin{figure}[b]
\includegraphics[width=0.48\linewidth]{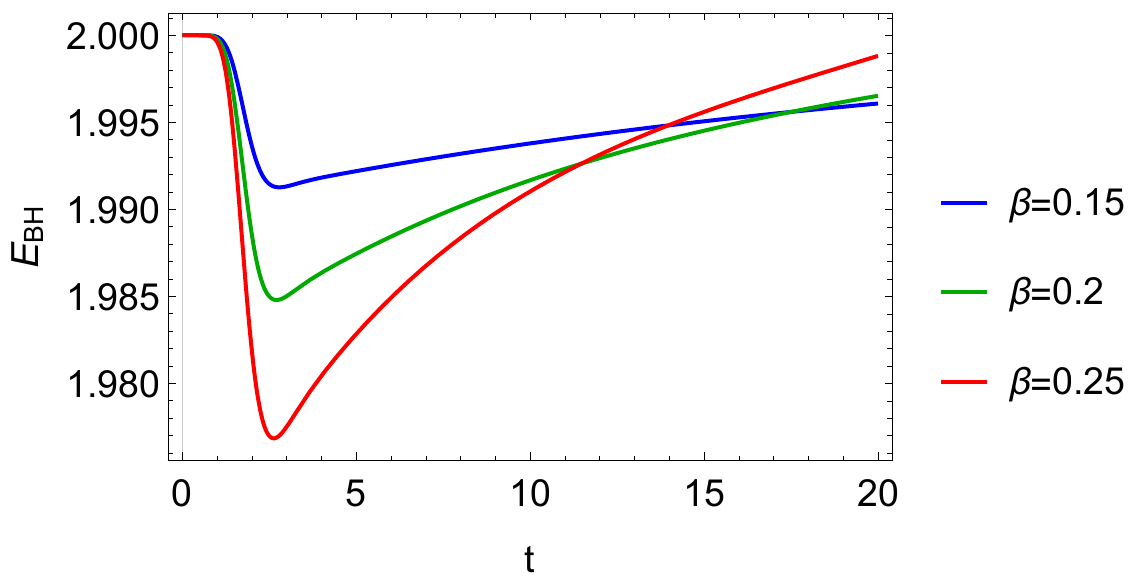}
\includegraphics[width=0.48\linewidth]{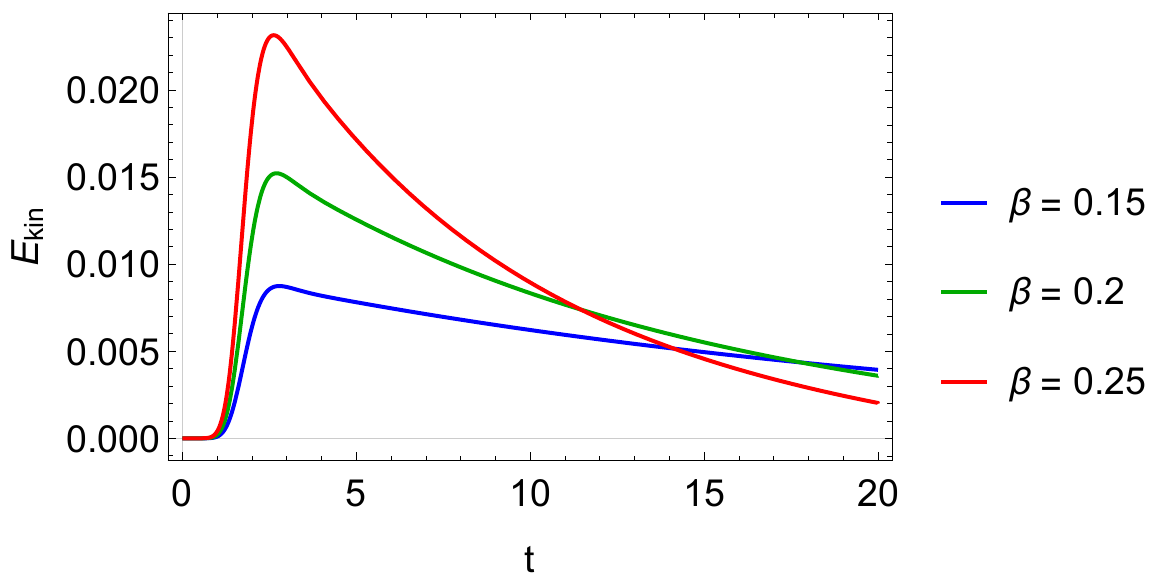}
\caption{Evolution of the black hole energy and the kinetic energy in the left and right panel, respectively, for a variety of couplings.}\label{fig:evol}
    \end{figure}

In Fig.~\ref{fig:homo-qnm}, there is also a conspicious unstable mode in the upper half plane. However, the full nonlinear simulations show that the apparently unstable mode indicates that energy is driven from the holographic sector to the boundary scalar at early times, as can be seen in Fig.~\ref{fig:evol}. 
Eventually the irreversible transfer of energy back to the holographic sector at later times is driven by the quasi-hydro mode. As such, the total system has no instability!

\begin{figure}
\includegraphics[width=0.48\textwidth]{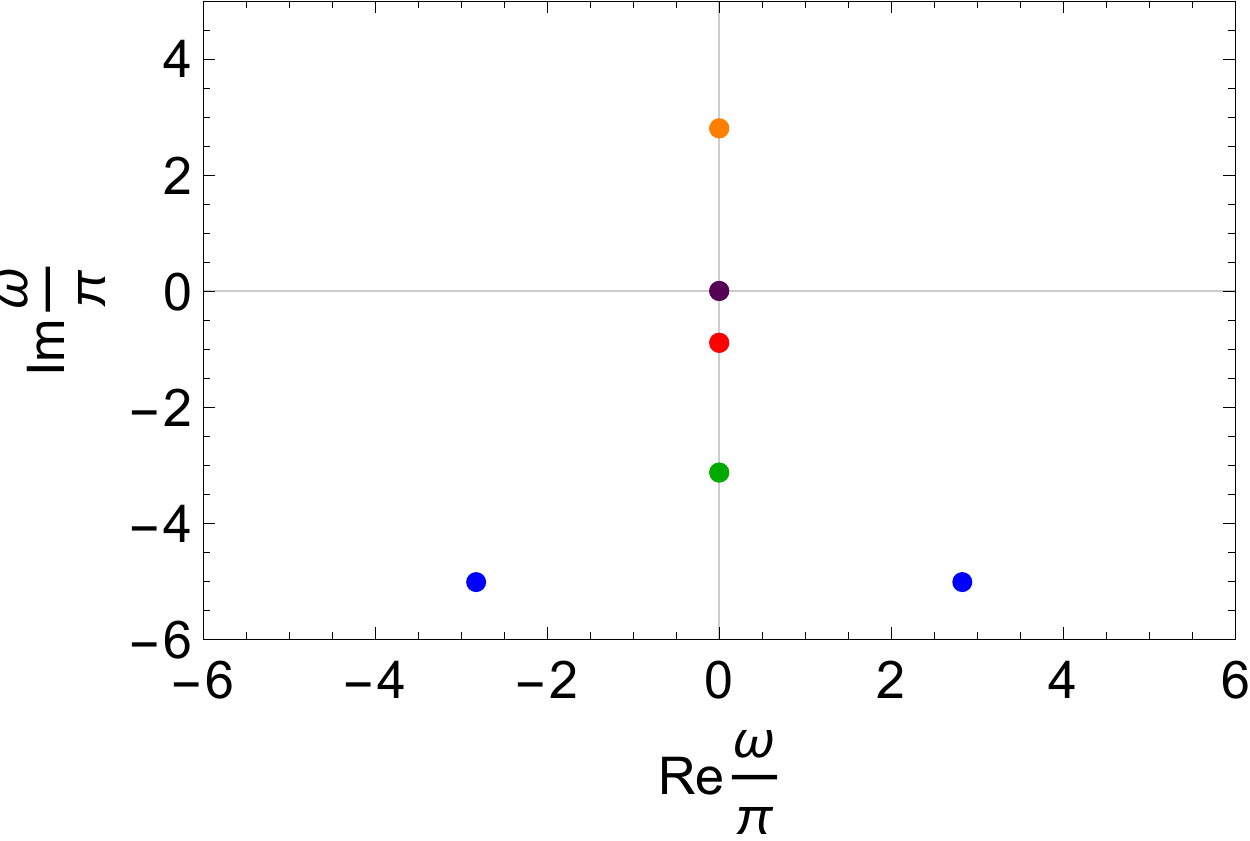}
\includegraphics[width=0.48\textwidth]{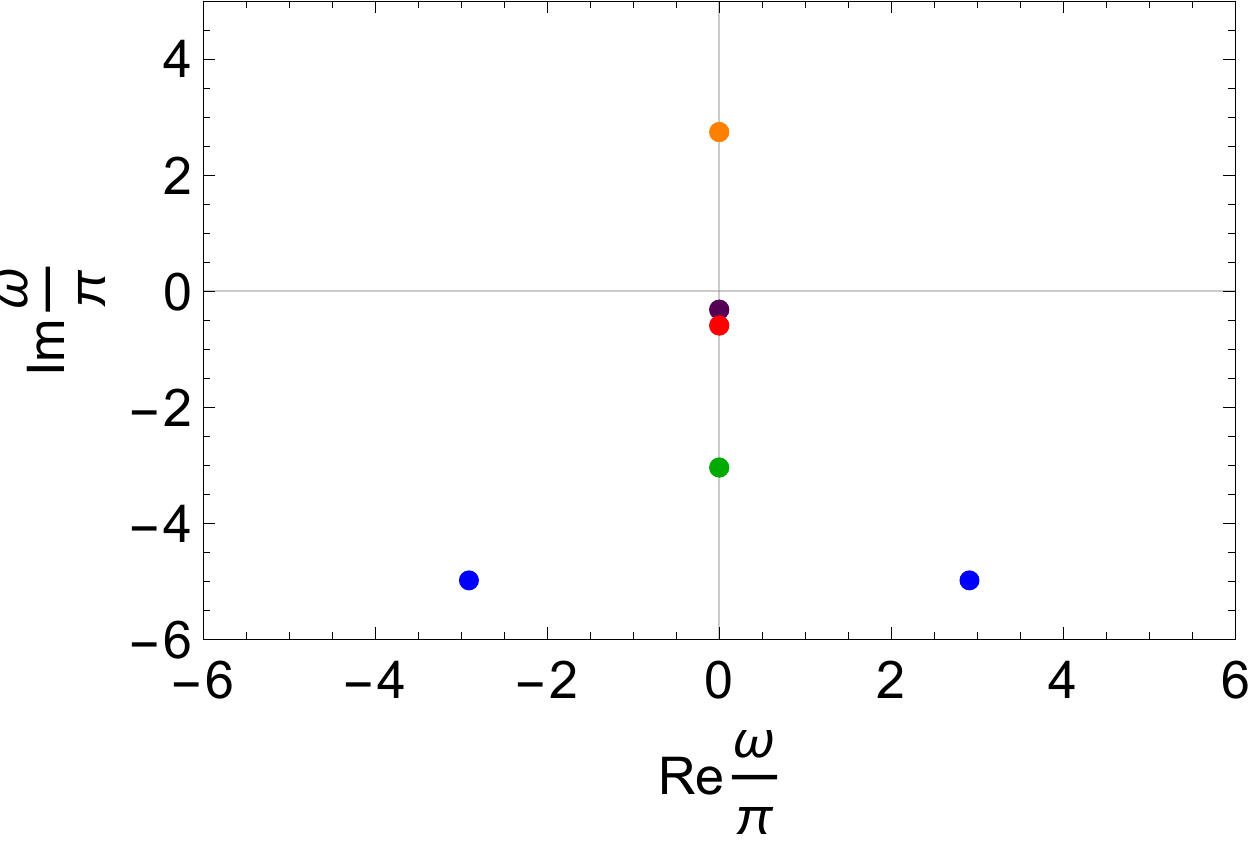}\\
{\includegraphics[width=0.48\textwidth]{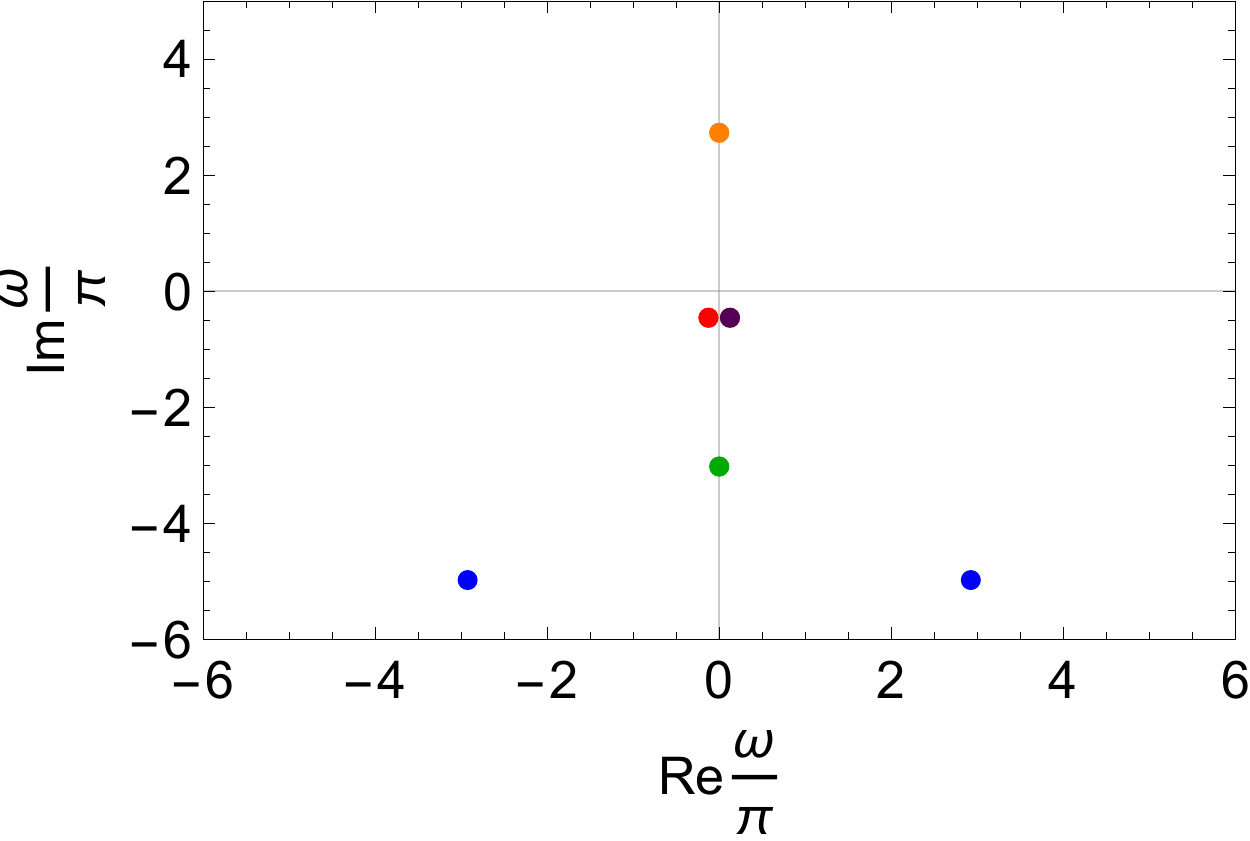}}
{\includegraphics[width=0.48\textwidth]{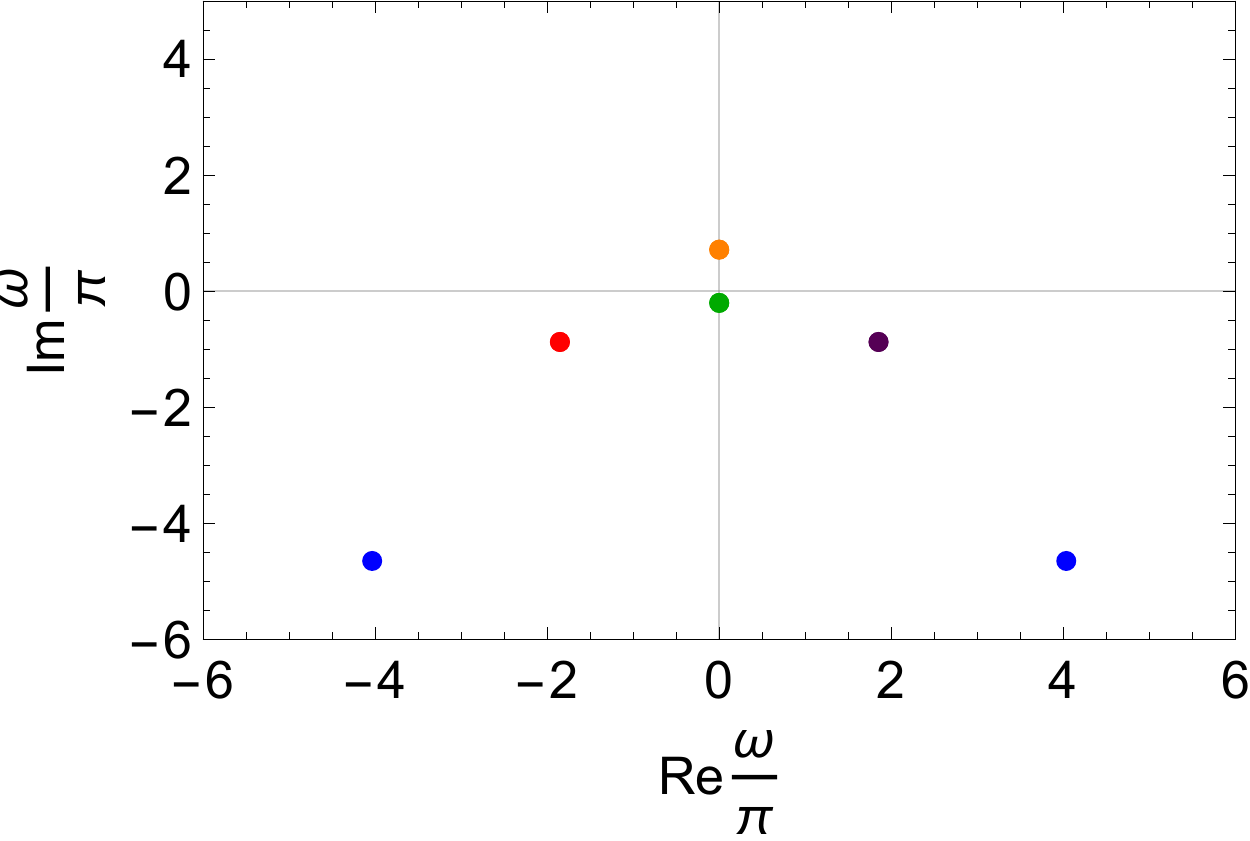}}\\
{\includegraphics[width=0.48\textwidth]{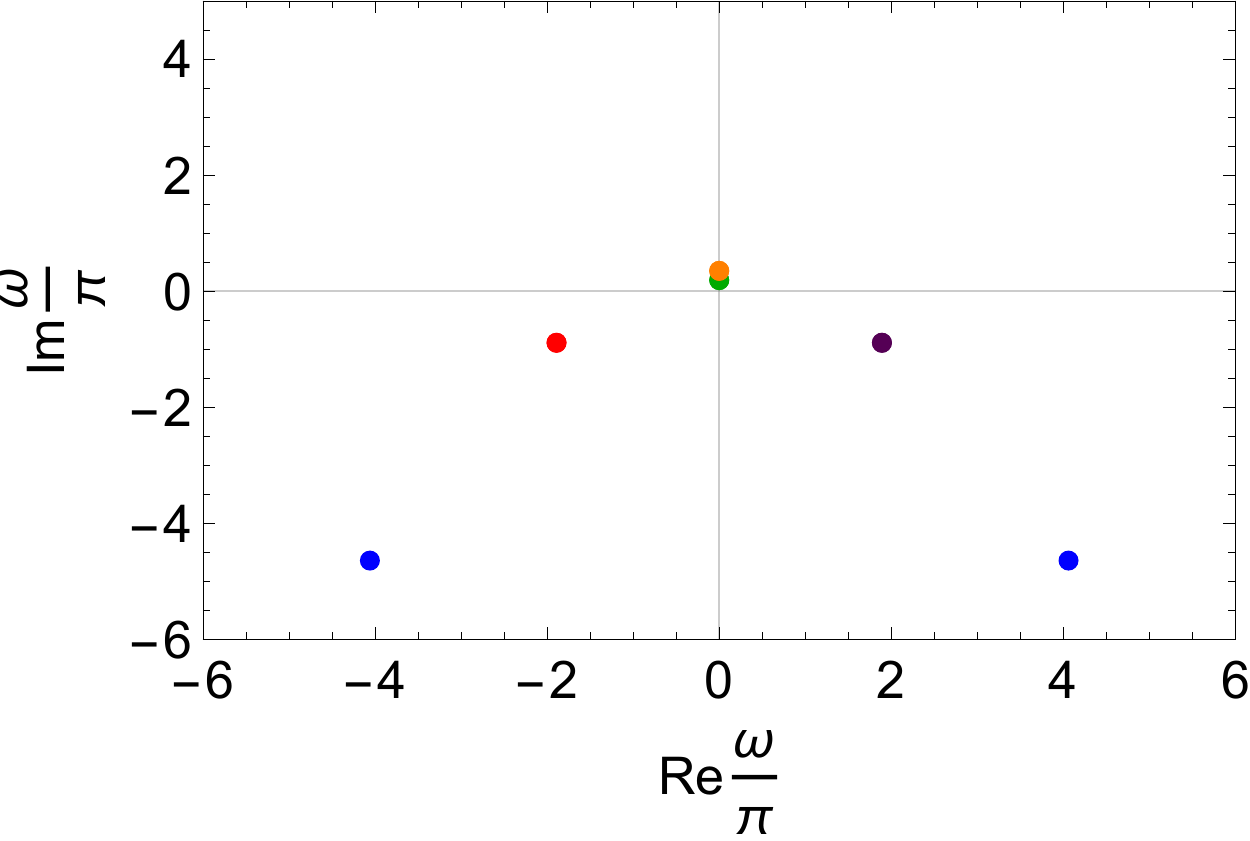}}
{\includegraphics[width=0.48\textwidth]{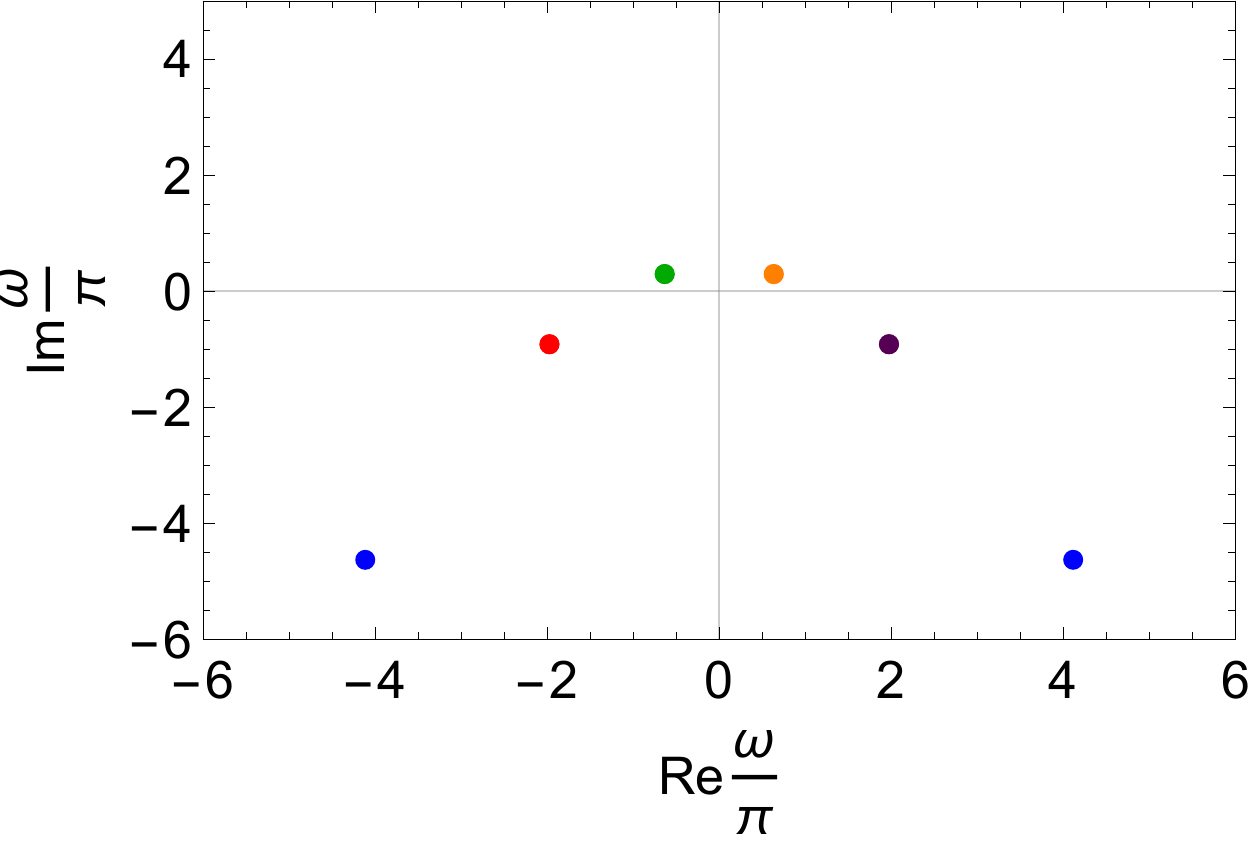}}
\caption{The first 6 poles in the complex plane for $k=\{0, 0.554\pi,0.6\pi,2.4\pi,2.4334\pi,2.54\pi \}$ (from top left to bottom right), for $\beta\sqrt{T}=0.35$. We distinguish the different poles via the following colors \textcolor{purple}{diffusion pole}, \textcolor{red}{quasi-hydro pole}, \textcolor{orange}{transient unstable mode},
\textcolor{green}{semiholographic pole} and \textcolor{blue}{holographic QNM}.}\label{fig:poles}
    \end{figure}

We now turn our attention to the inhomogeneous case. In particular, we examine the motion of the lowest lying poles for a demonstrative value of coupling, as can be seen in Fig.~\ref{fig:poles}. The first and last plot represent small and large $k$ behavior. The figures in between indicate the two collisions of poles that occur for intermediate momenta.  All collisions occur on the imaginary axis, with the poles acquiring a real part after the collision. The first collision occurs between the diffusion pole and the quasi-hydro pole for $k\sim 0.554\pi$, followed by a collision of the transient unstable mode (discussed above) and the semiholographic pole for $k\sim 2.433 \pi$. The two typical holographic poles (and the rest of the "Christmas tree" poles deeper in the complex plane) do not appreciably change as a function of momentum.

\begin{figure}
\includegraphics[width=0.49\textwidth]{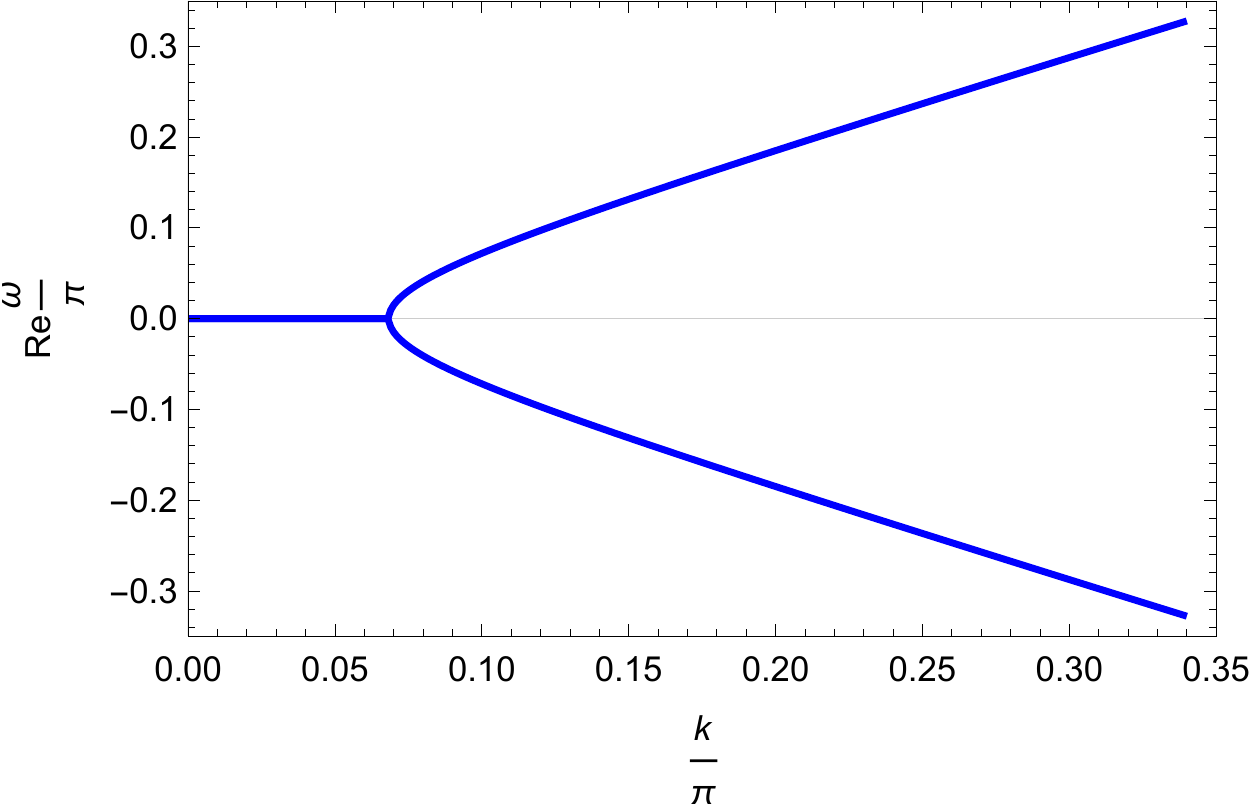}
\includegraphics[width=0.49\textwidth]{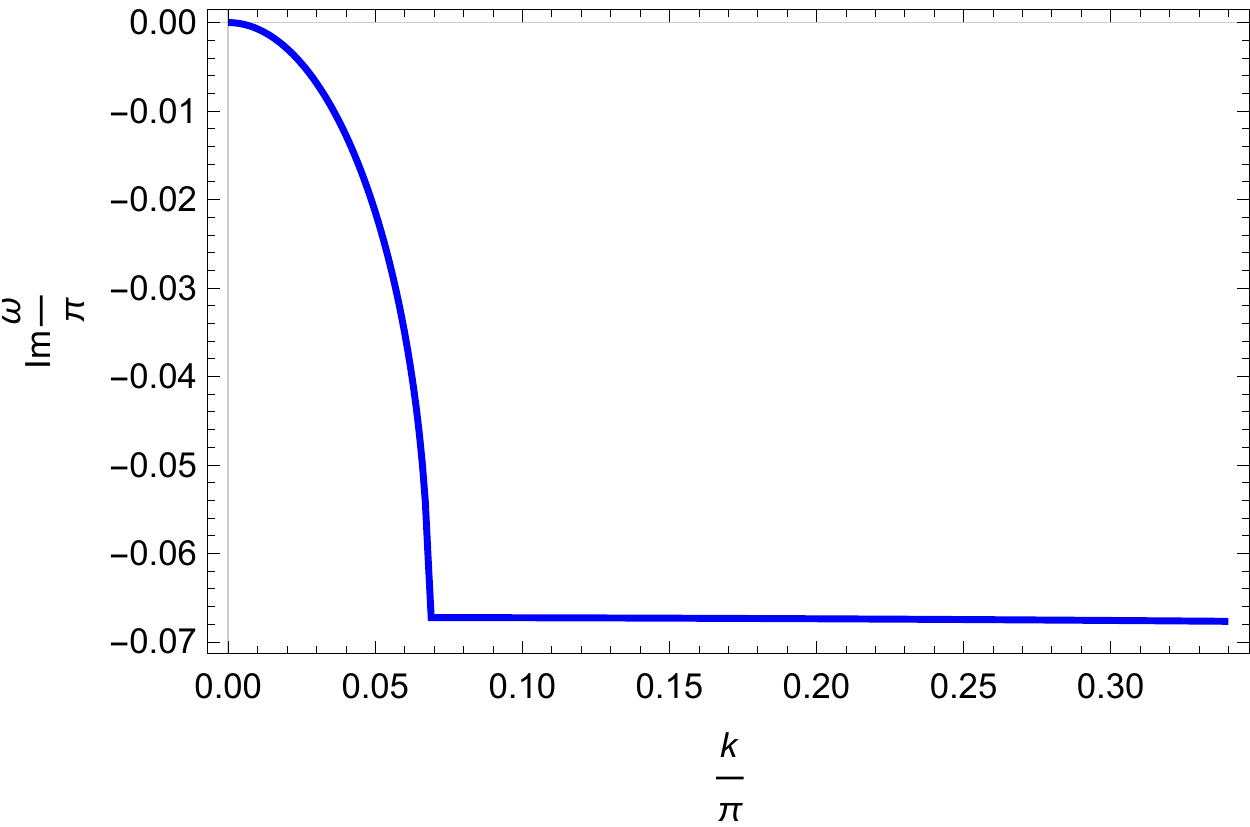}
\caption{The formation of the k-gap, shown for the particular case $\beta\sqrt{T}=0.15$.}\label{fig:kgap}
  \end{figure}

Finally, we also point out that in our model, we observe the formation of a $k$-gap as seen in Fig.~\ref{fig:kgap}, which is characteristic of systems with a diffusive to propagating mode crossover.
This has been observed in quasi-hydrodynamic frameworks \cite{Grozdanov:2018fic}, where the $k-$gap can be related to a softly broken global symmetry.  Here, global shift symmetry of the theory is 
\begin{align}
{\chi}\rightarrow{ \chi}+{\chi_0}, \quad {\Phi}\rightarrow  {\Phi}-\beta { \chi_0}
\end{align}
The k-gap has been observed experimentally in the dispersion of transverse sound-like excitations of gallium \cite{Khusnutdinoff_2020}.

\section{Outlook}\label{sec:outlook}

The analytic structure of the QGP is rich. Due to the presence of phase transitions, such as the chiral phase transition and the one due to strong/weak coupling dynamics, we should expect to see colliding poles in the complex plane of theories relevant to the QGP. In the hybrid model outlined Sec.~\ref{sec:semi}, an immediate follow-up includes understanding the interaction of the quasinormal mode spectrum of a holographic theory with Israel-Stewart hydrodynamics \cite{mondkar:tbp}. Another important direction is to see the interplay of QNM poles with branch cuts typical to kinetic theory, especially in the vicinity of a phase transition.

\section*{Acknowledgments}
The author is supported by the Austrian Science Fund (FWF), project no. J4406.

\bibliography{conf-bib}
\bibliographystyle{woc}

\end{document}